\documentclass[conference,letterpaper]{IEEEtran}
\usepackage[T1]{fontenc}
\usepackage{amsmath,amssymb,amsthm,mathtools}
\usepackage{booktabs,array,enumitem}
\usepackage{cite}
\usepackage[hidelinks]{hyperref}
\usepackage{microtype}
\usepackage{flushend,caption}
\usepackage[utf8]{inputenc} 
\usepackage[T1]{fontenc}
\usepackage{ifthen}
\usepackage{cite}

\newtheorem{theorem}{Theorem}
\newtheorem{corollary}{Corollary}

\newtheorem{remark}{Remark}
\newtheorem{lemma}{Lemma}

\newcommand{\F}{\mathbb F}

\newcommand{\one}{\mathbf 1}

\newcommand{\GL}{\operatorname{GL}}
\newcommand{\ket}[1]{\lvert #1\rangle}

\newenvironment{Proof}[1]{\medskip\par\noindent{\bf Proof:\,}\,#1}{{\mbox{\,$\blacksquare$}\par}}

\IEEEoverridecommandlockouts
\allowdisplaybreaks

\begin{document}

\title{Quantum Private Information Retrieval on Forests}

\author{Mohamed Nomeir  \qquad Sennur Ulukus\\
 Department of Electrical and Computer Engineering \\
 University of Maryland, College Park, MD 20742 \\
 \textit{mnomeir@umd.edu} \qquad  \textit{ulukus@umd.edu}}

\maketitle

\begin{abstract}
We consider the problem of graph private information retrieval (PIR) in the quantum setting. We focus on a specific type of graphs, namely the forest graphs. A forest graph is a union of disjoint trees. We consider the quantum setting where all the servers share an entangled state. Each server is allowed to send only one qudit per retrieval round. We design a scheme for the forest graphs that achieves the rate of $\frac{3}{N}$. The scheme uses a GHZ state as the initial entangled state between the servers prior to the scheme initiation.  We compare our scheme with the only work done in the literature that considered the quantum setting of graph PIR. We show that our scheme is more efficient in this setting. We derive the quantum capacity for the case of $G=3K_2$, which is equal to $4/6$. $G=3K_2$ is a special case of the forest graph. We design a scheme that is specialized for this graph, achieving a rate equal to $4/6$ and provide the corresponding matching upper bound, establishing it as the capacity.
\end{abstract}

\section{Introduction}\label{sec:intro}

Private information retrieval (PIR) has been introduced in \cite{chor}. The PIR system consists of $N$ servers with $K$ messages replicated in their storage. The user wants to retrieve one of the $K$ messages without revealing the required message index to any of the servers. The main metric for the efficiency in the PIR problem is the rate, which is the ratio between the required message length and the total number of downloaded symbols. The capacity is defined as the highest possible achievable rate. In \cite{c_pir}, it was shown that the capacity of such a setting with $N$ servers and $K$ replicated messages is $C= \left(1+\frac{1}{N}+\ldots+\frac{1}{N^{K-1}}\right)^{-1}$. Since then, many variants have appeared with different threat models, see e.g., \cite{c_TESPIR,ChaoTian,csa,byzantine_tpir,wang_spir,ulukusPIRLC,Salim_CodedPIR,star_graph,AG_2,arbit_pir_conf,arbit_pir_jour,local_pir_conf,local_pir_jour,nomeirasymmetric,Nomeir_symmetric_byzantine}. 

Recently, quantum communications has gained significant focus in the research community due to their potential to provide higher gains in general compared to their classical counterparts. It was shown in \cite{super_dense} that a qubit can carry two bits of information from the transmitter to the receiver if they share an entangled state prior to the scheme initiation. This was coined as super-dense coding gain. Another variant for multi-transmitter communication systems was developed that analyzes the setting with the transmitters sharing an entangled state without sharing with the receiver. In this setting it was shown that at most $N$ bits can be recovered from $N$ entangled transmitters due to Holevo's bound \cite{nielsen-chuang}. However, most recently, it was shown that for specific applications the rate of the quantum setting can achieve super-dense coding gains in some cases  \cite{Hu2025, yao_inverted, yao_Qsum_arbitrary_input_entanglemets, quantum_dist_storage_lei}. In addition, it was shown that for the quantum version of PIR (QPIR), the super-dense coding gain is achievable for certain regimes \cite{jafar_quantum_unresponsive, qtpir, qpir_colluding_mdscoded, byzantine_1, Byzantine_journal, our_quantum_first, our_journal}.

Due to the abundance of data, the focus in PIR research has shifted towards relaxing the full replication requirement. One of the ways of doing this is through graph PIR. In graph PIR, the nodes of the graph represent the servers and the hyper-edges represent the messages. Thus, a message is available at a subset of servers if the edge representing that message is incident to the nodes corresponding to these servers. Although the formulation of the graph version of PIR is straightforward, it remains one of the most challenging problems in the PIR literature. Many works have provided general upper bounds or specific capacities for certain graphs \cite{gePIR, graphbased_pir, star_graph, shreya_graph_1, shreya_graph_4, shreya_graph_5}. 

Most recently, \cite{Liu2025} provided a scheme for any arbitrary graph-based PIR in the classical setting. The scheme designed integrates well with the quantum setting as mentioned briefly in their abstract and a remark, thus achieving super-dense coding gain. Although the scheme in \cite{Liu2025} performs well in most quantum PIR settings, it can be inefficient in some specific scenarios, due to the generality of the scheme. To the best of our knowledge, \cite{Liu2025} is the only paper that mentions the quantum version of PIR over graphs. In this paper, we focus on one family of graph PIR, namely, forest graphs and analyze its quantum setting. We design a scheme for general forest graphs that achieves the rate $R_Q = 3/N$. We then focus on the matching graph $3K_2$, where there are 3 messages, 6 servers and each message is stored in exactly two servers. We design a specialized scheme for this graph that has a rate $R_Q = \frac{4}{6}$. We show that the achieved rate for $G=3K_2$ is the highest possible rate, thus resolving the capacity for this graph.

The rest of the paper is organized as follows. Section~\ref{sec:model} provides the problem formulation and the background needed for the $N$-sum box framework. In Section \ref{sec:main}, we provide the main results and comparisons with both the classical and quantum PIR results available in the literature. In Section~\ref{sec:scheme}, we provide the scheme for the general forest graphs. We provide an illustrative example in Section~\ref{sec:example}. Finally, in Section~\ref{sec:3K_2}, we provide the achievable scheme and the upper-bound derivation for $G=3K_2$, establishing the capacity for this case. 

\textit{Notations:} The symbol $\mathbb{F}_p$ is used to denote the finite field with $p$ elements, where $p$ is a prime. The set $\{1,2,\ldots,N\}$ is denoted by $[N]$ and the set $\{a,a+1,\ldots,b\}$ is denoted by $[a:b]$, where $a$ and $b$ are positive integers with $b > a$. GL$_m(\mathbb{F}_p)$ is the group of the generalized linear operators of size $m \times m$ with elements from $\mathbb{F}_p$. $\one_N$ is the all-ones column vector with length $N$. $K_N$ is used to denote the complete graph with $N$ nodes. We use $\sqcup$ to denote the union of disjoint sets. We denote the rate of the extended quantum scheme based on the work by Liu-Jia-Jia \cite{Liu2025} as $R_{\mathrm{LJJ,Q}}$.

\section{Problem Formulation}\label{sec:model}
Let $G=(V,E)$ be a simple graph where the vertices $V$ represent the servers and the edges $E$ represent the messages. Thus, an edge $e \in E$ that is incident to the two servers $u,v$ represents the presence of the message $W_e$ in both the $u$th and the $v$th server. We denote the messages as $W_{(u,v)}$ when necessary. Note that the number of servers is $N=|V|$ and the number of messages is $K = |E|$. Each message $W_{e}$ has $L$ symbols that are generated uniformly at random from $\mathbb{F}_p^L$, i.e.,
\begin{align}
    H(W_{[K]})=\sum_{i \in [K]}H(W_i)=\sum_{e \in E}H(W_e) = KL. 
\end{align}
A user who chooses $\Theta = \theta$ uniformly at random from $[K]$ wants to retrieve $W_{\theta}$ without revealing the required message index $\theta$ to any server. To do that, the user sends queries $Q_n^{[\theta]}$ to the $n$th server. Since the user has no prior knowledge about the message contents, we have
\begin{align}
    I(W_{[K]}; \Theta, Q_{[N]}^{[\Theta]})=0.
\end{align}
In addition, to ensure user privacy
\begin{align}\label{eq:privacy}
    I(\Theta; \mathcal{W}_n, Q_{n}^{[\Theta]})=0, \quad n \in [N], 
\end{align}
where $\mathcal{W}_n$ is the storage at the $n$th server. Once the $n$th server receives the query $Q_n^{[\theta]}$, it generates an answer $A_n^{[\theta]}$ that is a deterministic function of the storage and the received query, i.e.,
\begin{align}
    H(A_n^{[\Theta]}|\mathcal{W}_n, Q_n^{[\Theta]})= 0, \quad n \in [N].
\end{align}
In the quantum setting, we assume that the servers share an entangled state represented by the density matrix $\rho$ in the Hilbert space $\mathcal{H}$, with $\log\mathrm{dim}(\mathcal{H}) = N$. The $n$th server encodes the answer over its share of the entangled state $\rho_n = \mathrm{tr}_{i \in [N]\setminus n}(\rho)$ and transmits to the user over a noise-free quantum channel
\begin{align}
    \rho_n' = \mathrm{Enc}_n(\rho_n, A_n^{[\Theta]}).
\end{align}
Once the user receives all the quantum states $\rho_{[N]}'$ and applies a POVM to get the classical information $Y_{\Theta}$, the user must be able to decode the required message, i.e.,
\begin{align}
    H(W_{\Theta}|Y_{\Theta}, Q_{[N]}^{[\Theta]}) = 0.
\end{align}
The rate is defined as the ratio between the size of the required message and the number of received qudits. Since there are $N$ received qudits and the message lengths are equal $L$, the rate is given by $R_Q= \frac{L}{N}$. Thus, the goal is to maximize $L$ in order to increase the rate.

In this paper, we focus on forest graphs only.
We use standard graph terminology \cite{vanLintWilson2001}.
A forest is a disjoint union of trees. A nontrivial tree has at least
two vertices. Thus, a forest graph $G$ can be written as a union of disjoint trees, i.e.,
\begin{align}\label{eq:forest}
 G=T_1\sqcup\cdots\sqcup T_c,\qquad c\ge2,
 \end{align}
with $|V(T_j)|\ge2$, $j \in [c]$. Thus, there are no isolated vertices, $N\ge4$, and $K=N-c\ge2$.

Prior to the scheme initiation, we orient every edge publicly and fix this orientation before retrieval, i.e.,
the notation $e=(u_e,v_e)$ specifies its tail and head. 

We make use of the $N$-sum box abstraction in \cite{nsumbox}. The main theorem that we make use of is stated as follows.

\begin{theorem}[Theorem~1  in \cite{nsumbox}]\label{feasible_nsum_1}
    Let ${G}$ be a $2N\times N$ SSO matrix, i.e., ${G^tJG}=0$, where ${J} = \begin{bmatrix}
        0 &{I}_N\\
        -{I}_N&0
    \end{bmatrix}$, and let ${H}$ be a $2N\times N$ matrix with full column rank. If $[{G} ~ {H}]$ is a full-rank matrix, then $M = [0_N ~ {I}_N ] [{G} ~ {H} ]^{-1}$ is a feasible $N$-sum box transfer matrix. 
\end{theorem}

\section{Main Results}\label{sec:main}

\begin{theorem}\label{thm:main}
For every forest satisfying \eqref{eq:forest} and every prime $p>N$,
there is a one-round, zero-error PIR scheme with one fixed $N$-party GHZ state, and exactly one $p$-dimensional qudit transmitted by each server, that achieves a rate
\begin{align}\label{eq:mainrate}
R_Q=\frac{3}{N}.
\end{align}
\end{theorem}

The scheme is given in Section~\ref{sec:scheme}.

\begin{theorem}\label{thm:3K_2}
    For $G = 3K_2$, the quantum PIR capacity is given by
    \begin{align}
        C_Q = \frac{4}{6}.
    \end{align}
\end{theorem}
The scheme and the converse are given in Section~\ref{sec:3K_2}.
\begin{corollary}[\bf Comparison with the graph-based quantum extension in \cite{Liu2025}]
\label{cor:jia}
Specialize \cite[Thm.~1, Eqn.~(15), and Remark~3]{Liu2025}
to two replicas, uncoded replication $\eta_m=1$, $X=0$, and $T=1$.
Its rate is $R_{\mathrm{LJJ,Q}}=\frac2N$. Thus, for every graph in Theorem~\ref{thm:main},
\begin{align}\label{eq:gain}
 \frac{R_Q}{R_{\mathrm{LJJ,Q}}}=\frac32.
\end{align}
\end{corollary}

\begin{Proof}
Using the same notation as in the original paper \cite{Liu2025}, for a message group stored at the two endpoints $\{u,v\}$, the
one-element-subset constraints in the cited achievable region require
both classical normalized downloads to satisfy $d_u\ge1$ and $d_v\ge1$.
Every vertex is incident to an edge, thus the minimum total normalized
classical download within this region is $N$. \cite[Remark~3]{Liu2025} gives the corresponding quantum normalized download $N/2$.
\end{Proof}

\begin{corollary}
\label{cor:classical}
For a matching $G=mK_2$ with $m\ge2$ and $N=2m$, classical PIR capacity is
$C_{\rm C}(G)=1/m=2/N$. Theorem~\ref{thm:main} achieves the quantum
rate $3/N$, exceeding that classical capacity by a factor of $3/2$. In addition, for $m=3$, Theorem~\ref{thm:3K_2} achieves super-dense coding gain, i.e., gain by a factor of $2$.
\end{corollary}

\begin{Proof}
Label the messages (edges) of $G=mK_2$ as $e_i=\{2i-1,2i\},~ i\in[m]$, and let $W_i$ be the file stored at the two endpoints of $e_i$. Thus, servers $2i-1$ and $2i$ store $W_i$. In addition, no other server
stores this file. 

\emph{Converse:}
Consider an arbitrary classical PIR scheme.
From the individual server privacy, we know that $H(A_n^{[i_1]})=H(A_n^{[i_2]})= \ldots=H(A_n^{[i_K]})$. Thus, we can define
\begin{align}
    d_n = H(A_n^{[i]}),\qquad n\in[N],
\end{align}
using any desired index $i$. In particular, the same vector
$(d_1,\ldots,d_N)$ describes the server-answer entropies for
every desired file.

We now apply the argument of
\cite[Lemma~3]{Sadeh2021} to edge $e_i$.
Let $\mathcal P_i=\{2i-1,2i\}$, and $W_{-i}=(W_j)_{j\in[m]\setminus\{i\}}$. Independence of the files and queries gives
\begin{align}\label{eq:matching_file_entropy}
    H(W_i\mid W_{-i},Q_{[N]}^{[i]})=L.
\end{align}
Every answer outside $\mathcal P_i$ is a deterministic
function of its local query and a file contained in $W_{-i}$.
Therefore,
\begin{align}\label{eq:matching_outside_answers}
    H\!\left(
      A_{[N]\setminus\mathcal P_i}^{[i]}
      \mid W_{-i},Q_{[N]}^{[i]}
    \right)=0.
\end{align}
Thus, given $W_{-i}$ and $Q_{[N]}^{[i]}$, the user can reconstruct
all answers outside $\mathcal P_i$. Zero-error correctness
then implies
\begin{align}\label{eq:matching_pair_decodes}
    H\!\left(
      W_i\mid A_{\mathcal P_i}^{[i]},
      W_{-i},Q_{[N]}^{[i]}
    \right)=0.
\end{align}
Combining \eqref{eq:matching_file_entropy} and
\eqref{eq:matching_pair_decodes}, we obtain
\begin{align}
    L
    &= I\!\left(
         W_i;A_{\mathcal P_i}^{[i]}
         \mid W_{-i},Q_{[N]}^{[i]}
       \right) \\ 
       &\le H\!\left(
         A_{\mathcal P_i}^{[i]}
         \mid W_{-i},Q_{[N]}^{[i]}
       \right) \\
    &\le H(A_{2i-1}^{[i]})+H(A_{2i}^{[i]})\\ &=d_{2i-1}+d_{2i}.
    \label{eq:matching_edge_constraint}
\end{align}

Although equality in \eqref{eq:matching_edge_constraint} is derived by
considering retrieval of $W_i$, privacy expresses every such
inequality using the same quantities $d_n$. Moreover, the
$m$ edges are vertex-disjoint and cover all $N=2m$ servers.
Summing over $i\in[m]$ gives
\begin{align}
    \sum_{n=1}^{N}d_n
    =\sum_{i=1}^{m}(d_{2i-1}+d_{2i})
    \ge mL.
\end{align}
Hence, every classical PIR scheme satisfies
\begin{align}
    R_{\rm C}
    =\frac{L}{\sum_{n=1}^{N}d_n}
    \le\frac1m.
\end{align}
This is precisely the matching-number converse
$C_{\rm C}(G)\le1/\nu(G)$ in
\cite{Sadeh2021}, specialized to $\nu(G)=m$.

\emph{Achievability:}
Choose an even file length $L$ and split every file into
two equal-length parts, $W_i=(W_{i,\mathrm{L}},W_{i,\mathrm{R}})$, with each having length $\frac{L}{2}$. For every desired index, send the same fixed queries:
server $2i-1$ returns $W_{i,\mathrm{L}}$, and server $2i$
returns $W_{i,\mathrm{R}}$, for every $i\in[m]$. The user consequently recovers all $m$ files, including the
desired file. Privacy holds because the queries do not
depend on the desired index. Each server returns $L/2$
symbols, therefore the achieved rate is $1/m$, showing
\begin{align}
    C_{\rm C}(mK_2)=\frac1m=\frac2N,
\end{align}
and completing the proof.
\end{Proof}
\begin{remark}
    Since the path graph $P_N$ is a tree (note that $K_2$ is also a path), we can have a disjoint union of paths as a forest graph, i.e., $G = mP_N$ is a forest. It was shown in \cite{shreya_graph_3} that the capacity for this setting is $C(mP_N) = \left(\sum_{i \in [m]}C(P_N)^{-1}\right)^{-1} = \frac{2}{mN}$, which coincides with the above proof for $mK_2$. 
\end{remark}

\begin{remark}
    A star graph with one main hub and $N$ leaves, denoted $K_{1,N}$, is also considered as a tree graph. In \cite{star_graph}, it was shown that $C(K_{1,4}) = \frac{5}{12}$. Thus, for $G = 2 K_{1,4}$, we have $C(G) = \frac{5}{24}$. In comparison with Theorem~\ref{thm:main}, the rate gain is $R_Q/R_C = 1.44$.
\end{remark}
\section{Scheme for Theorem~\ref{thm:main}}\label{sec:scheme}
We use the generalized Pauli operators
\begin{align}
 X(b)\ket{t}&=\ket{t+b},\\
 Z(a)\ket{t}&=\omega^{at}\ket{t},
\end{align}
where $ \omega=\exp(2\pi\mathrm i/p)$ and $p$ is a prime \cite{nielsen-chuang}.
The following lemma provides the main quantum encoding and measurements used in the forest graph. We show the explicit $N$-sum box abstraction \cite{nsumbox} afterwards.
\begin{lemma}\label{lem:ghz}
Suppose server $n$ knows two scalars $a_n,b_n\in\F_p$ and has one qudit from the GHZ state represented by
\begin{align}\label{eq:ghz}
 \ket{\operatorname{GHZ}_{N,p}}
 =\frac1{\sqrt p}\sum_{t\in\F_p}\ket{t}^{\otimes N}.
\end{align} 
The user can recover
\begin{align}\label{eq:outputs}
 S=\sum_{n=1}^N a_n,\qquad
 \Delta_i=b_i-b_N,\quad i\in[N-1],
\end{align}
by receiving exactly one qudit from each server.
\end{lemma}

\begin{Proof}
Server $n$ applies $X(b_n)Z(a_n)$ to its own qudit and sends that
qudit to the user. The received pure state is
\begin{align}
 \frac1{\sqrt p}\sum_t
 \omega^{St}\ket{t+b_1,\ldots,t+b_N}.
\end{align}
Set $z=t+b_N$ and $\Delta_i=b_i-b_N$. Up to a global phase
$\omega^{-Sb_N}$, this is
\begin{align}\label{eq:ghzbasis}
 \ket{\Psi_{S,\Delta}}=\frac1{\sqrt p}\sum_z
 \omega^{Sz}\ket{z+\Delta_1,\ldots,z+\Delta_{N-1},z}.
\end{align}
Different $\Delta = (\Delta_1, \ldots, \Delta_{N-1})$ values give disjoint computational basis supports.
For a fixed $\Delta$, the inner product for two phase labels $S,S'$
is $p^{-1}\sum_z\omega^{(S-S')z}$, which is one if $S=S'$ and zero
otherwise. Thus, the $p^N$ states in \eqref{eq:ghzbasis} form an
orthonormal basis for the $N$-qudit space. Measuring in that fixed basis
recovers every label in \eqref{eq:outputs} with certainty.
\end{Proof}

For mapping this to the $N$-sum box abstraction \cite{nsumbox}, let the input order be
$(a_1,\ldots,a_N,b_1,\ldots,b_N)$, where the $i$th transmitter has the pair $(a_i,b_i)$, the transfer matrix is
\begin{align}\label{eq:matrix}
 M_N=\begin{bmatrix}
       \one_N^\top&0\\
       0&D_N
     \end{bmatrix},
 \end{align}
 where $D_N=[I_{N-1}, -\one_{N-1}]$.
It has rank $N$. To see that it is a valid $N$-sum box, we make use of Theorem~\ref{feasible_nsum_1}. Define
\begin{align}
e_N&=\begin{bmatrix}0\\ \vdots\\0\\1\end{bmatrix},\quad
E_N=\begin{bmatrix}I_{N-1}\\0_{1\times(N-1)}\end{bmatrix}.
\end{align}
Then, $G$ and $H$ are given as follows
\begin{align}
G&=
\begin{bmatrix}
0_{N\times1}&D_N^\top\\
-\mathbf1_N&0_{N\times(N-1)}
\end{bmatrix},\\    
H&=
\begin{bmatrix}
e_N&0_{N\times(N-1)}\\
0_{N\times1}&E_N
\end{bmatrix}.
\end{align}
Noe that $G$ has rank $N$, and 
\begin{align}
    G^t J_N G
=
\begin{bmatrix}
0&\mathbf1_N^\top D_N^\top\\
-D_N\mathbf1_N&0
\end{bmatrix}
=0,
\end{align}
because \(D_N\mathbf1_N=0\).

Moreover, $[G ~H]$ is invertible, with
\begin{align}
    \begin{bmatrix}
        G&H
    \end{bmatrix}^{-1}=\begin{bmatrix}
0_{1\times N}&-e_N^\top\\
E_N^\top&0_{(N-1)\times N}\\
\mathbf1_N^\top&0_{1\times N}\\
0_{(N-1)\times N}&D_N
\end{bmatrix}.
\end{align}
Thus, \eqref{eq:matrix}
is a valid $N$-sum box.

\subsection{Queries and  Encoding}
Independently for each edge $e$, the user samples the matrix $R_e \in \mathrm{GL}_3(\mathbb{F}_p)$ uniformly at random. It can be written as
\begin{align}
R_e=\begin{bmatrix}r_{e,1}\\r_{e,2}\\r_{e,3}\end{bmatrix}.
\end{align}
Write the transformed coordinates as
\begin{align}\label{eq:transformed}
 R_eW_e=[A_e,B_e,C_e]^t.
\end{align}
The user knows $R_e$ but does not know these coordinate values.
A server computes only the linear forms specified in its query.

Let the desired edge be $\theta=(u_\theta,v_\theta)$.
For every oriented edge $e=(u_e,v_e)$, the user sends the tail of the two row vectors 
\begin{align}\label{eq:tailquery}
 \begin{bmatrix}
     \alpha_{u_e,e}\\ \beta_{u_e,e}
 \end{bmatrix}=
 \begin{bmatrix}
     r_{e,1}\\r_{e,2}
 \end{bmatrix},
\end{align}
and sends its head the following two row vectors
\begin{align}\label{eq:headquery}
 \begin{bmatrix}
     \alpha_{v_e,e}\\
     \beta_{v_e,e}
 \end{bmatrix}
 \begin{cases}
  \begin{bmatrix}
    r_{e,3}\\r_{e,1}  
  \end{bmatrix},&e=\theta,\\
  \begin{bmatrix}
    -r_{e,1}\\
    -r_{e,2}  
  \end{bmatrix},&e\ne\theta.
 \end{cases}
\end{align}
Server $n$ receives just these two coefficient rows for each of its
stored files, then computes two scalars from its stored files
\begin{align}\label{eq:local}
 a_n=\sum_{e\ni n}\alpha_{n,e}W_e,\qquad
 b_n=\sum_{e\ni n}\beta_{n,e}W_e.
\end{align}
It applies $X(b_n)Z(a_n)$ to its single qudit of
\eqref{eq:ghz}.
The user applies Lemma~\ref{lem:ghz} and obtains \eqref{eq:outputs}.

For $e\ne\theta$, the contributions of its endpoints to $(a_n,b_n)$
are
\begin{align}\label{eq:cancelpairs}
 (A_e,B_e)\quad\text{and}\quad(-A_e,-B_e).
\end{align}
For $e=\theta$, they are
\begin{align}\label{eq:desiredpairs}
 (A_\theta,B_\theta)\quad\text{and}\quad(C_\theta,A_\theta).
\end{align}
Thus, $S=A_\theta+C_\theta$.
\subsection{Decoding}

It is clear that deleting the edge $\theta$ from $G$ splits its tree into two vertex sets
$\mathcal A_\theta$ and $\mathcal B_\theta$, containing
$u_\theta$ and $v_\theta$, respectively. The user chooses any other original
tree component, we can denote its vertex set by $\mathcal C$. Note that such a component always exists because $c\ge2$.
These three sets are known to the user since the storage graph $G$ is known.

\begin{lemma}\label{lem:cuts}
The inputs in \eqref{eq:local} satisfy
\begin{align}
 \sum_{n\in\mathcal C}b_n&=0,\label{eq:anchor}\\
 \sum_{n\in\mathcal A_\theta}b_n&=B_\theta,\label{eq:cutA}\\
 \sum_{n\in\mathcal B_\theta}b_n&=A_\theta.\label{eq:cutB}
\end{align}
\end{lemma}

\begin{Proof}
Every undesired edge internal to a vertex set contributes
$B_e-B_e=0$ by \eqref{eq:cancelpairs}.
No edge crosses an original component $\mathcal C$.
The only edge crossing either component of the tree with $\theta$
removed is $\theta$ itself: a second crossing edge would create a cycle.
Thus, the only uncancelled contribution in $\mathcal A_\theta$ is
$B_\theta$ at its tail, and the only one in $\mathcal B_\theta$ is
$A_\theta$ at its head.
\end{Proof}

Set $\Delta_N=0$ in addition to the measured differences, and let
$\zeta=b_N$. For every server,
\begin{align}\label{eq:shift}
 b_n=\Delta_n+\zeta.
\end{align}
Note that \eqref{eq:anchor} determines the unknown common shift
\begin{align}\label{eq:offset}
 \zeta=-|\mathcal C|^{-1}\sum_{n\in\mathcal C}\Delta_n.
\end{align}
The inverse exists because $1\le|\mathcal C|\le N<p$.
The user can now compute every $b_n$.

The three desired coordinates are recovered explicitly as
\begin{align}
 A_\theta&=\sum_{n\in\mathcal B_\theta}b_n,\label{eq:recoverA}\\
 B_\theta&=\sum_{n\in\mathcal A_\theta}b_n,\label{eq:recoverB}\\
 C_\theta&=S-A_\theta.\label{eq:recoverC}
\end{align}
Finally, $W_\theta=R_\theta^{-1}(A_\theta,B_\theta,C_\theta)^t$.

It is clear that privacy is guaranteed from each server since $R_e$ is generated uniformly at random and independently of $\Theta$. Thus, each of $[r_{e,1}^t,r_{e,2}^t]^t$, $[r_{e,3}^t,r_{e,1}^t]^t$ and $[-r_{e,1}^t,-r_{e,2}^t]^t$ is independent of $\Theta$ and has rank $2$.

\section{Illustrative Example}
\label{sec:example}

Let $G=K_2\sqcup K_2$, with $W_1$ at servers $1$, $2$ and $W_2$
at servers $3$, $4$. Every file has three independent $\F_p$ symbols,
where $p>4$. Let the user draw $R_1$, $R_2$ independently and uniformly from
$\GL_3(\F_p)$, and let
\begin{align}
 R_1W_1&=\begin{bmatrix}
     a\\b\\c
 \end{bmatrix}, \qquad R_2W_2=\begin{bmatrix}
     d\\e\\f
 \end{bmatrix}.
\end{align}
Every server receives only its coefficient rows. Table~\ref{tab:example} shows the symbols transmitted per server.

\begin{table}[h]
\centering
\begin{tabular}{|c|c|c|c|}
\hline
Server & Stored file & Desired $W_1$ & Desired $W_2$\\
\hline
$1$ & $W_1$ & $(a,b)$ & $(a,b)$\\

\hline
$2$ & $W_1$ & $(c,a)$ & $(-a,-b)$\\
\hline
$3$ & $W_2$ & $(d,e)$ & $(d,e)$\\
\hline
$4$ & $W_2$ & $(-d,-e)$ & $(f,d)$\\
\hline
\end{tabular}
\caption{Inputs $(a_n,b_n)$ for $G=K_2\sqcup K_2$.}
\label{tab:example}
\end{table}

The user receives the outputs of the transfer matrix
\begin{align}\label{eq:M4}
 M_4=\left(\begin{array}{cccc|cccc}
 1&1&1&1&0&0&0&0\\
 0&0&0&0&1&0&0&-1\\
 0&0&0&0&0&1&0&-1\\
 0&0&0&0&0&0&1&-1
 \end{array}\right).
\end{align}
Recall that this matrix and the shared GHZ state and the measurements provided in Lemma \ref{lem:ghz} are the same.

If $W_1$ is being retrieved, the measured labels are
\begin{align}
 (S,\Delta_1,\Delta_2,\Delta_3)
 =(a+c,\ b+e,\ a+e,\ 2e).
\end{align}
The user computes
$e=\Delta_3/2,~ a=\Delta_2-e,~
 b=\Delta_1-e,~ c=S-a$. The user then recovers $W_1=R_1^{-1}(a,b,c)^t$.

In the case of retrieving $W_2$, the measured labels become
\begin{align}
 (S,\Delta_1,\Delta_2,\Delta_3)
 =(d+f,\ b-d,\ -b-d,\ e-d).
\end{align}
Hence, $d=-\frac{\Delta_1+\Delta_2}{2},~
 e=\Delta_3+d,~ f=S-d$, and the user obtains $W_2=R_2^{-1}(d,e,f)^t$.

\section{Proof of Theorem~\ref{thm:3K_2}}\label{sec:3K_2}
We first choose the field size $p = 7$ and  appropriate $N$-sum box as follows. Define 
\begin{align}
    B &= \begin{bmatrix}
        0&1\\
        2&0
    \end{bmatrix}, \qquad
    \Gamma = \begin{bmatrix}
       0&I_2&B\\
       I_2&0&I_2\\
       B^t&I_2&0
    \end{bmatrix}.
\end{align}
Using the representation of the $N$-sum box matrix $M = [0_N~I_N][G~H]^{-1}$, we choose
\begin{align}
    G &= \begin{bmatrix}
        \Gamma\\
        I_6
    \end{bmatrix}, \qquad
    H = \begin{bmatrix}
        I_6\\
        0_6
    \end{bmatrix}.
\end{align}
It can be verified that $M$ satisfies the $N$-sum box required properties as follows
\begin{align}
    G^tJ_6G &= \begin{bmatrix}
        -\Gamma^t & I_6
    \end{bmatrix}\begin{bmatrix}
        I_6\\
        \Gamma
    \end{bmatrix}\\
    &= \Gamma^t - \Gamma = 0,
\end{align}
where $\Gamma$ is symmetric and recall that \begin{align}
    J_6 = \begin{bmatrix}
        0&I_6\\
        -I_6&0
    \end{bmatrix}.
\end{align} In addition
\begin{align}
    \begin{bmatrix}
        G & H
    \end{bmatrix}^{-1} = \begin{bmatrix}
        0&I_6\\
        I_6& \Gamma
    \end{bmatrix}.
\end{align}

Now, to design the queries, for each $e \in [3]$, generate $R_e \in \mathrm{GL}_4(\mathbb{F}_7)$ uniformly at random. Denote the $i$th row of $R_e$ as $r_{e,i}$ and define 
\begin{align}
    R_eW_e = [a_e,b_e,c_e,d_e]^t.
\end{align}
Table \ref{tab:queries_3K2} provides the queries sent to each server for each required message index. It is clear that privacy is guaranteed since the scalar multiplication/division is done over a finite field.
\begin{table}[h]
\centering
\begin{tabular}{|c|c|c|c|}
\hline
Server & Request $W_1$ & Request $W_2$ & Request $W_3$ \\
\hline
$S_1$
& $(r_{1,1},r_{1,3})$
& $(\frac12 r_{1,1},r_{1,2})$
& $(r_{1,1},r_{1,2})$ \\
\hline
$S_2$
& $(r_{1,2},r_{1,4})$
& $(2r_{1,2},\frac12 r_{1,1})$
& $(r_{1,2},\frac12 r_{1,1})$ \\
\hline
$S_3$
& $(\frac12 r_{2,2},r_{2,1})$
& $(r_{2,1},r_{2,3})$
& $(2r_{2,1},r_{2,2})$ \\
\hline
$S_4$
& $(r_{2,1},r_{2,2})$
& $(r_{2,2},r_{2,4})$
& $(\frac12 r_{2,2},2r_{2,1})$ \\
\hline
$S_5$
& $(r_{3,2},r_{3,1})$
& $(2r_{3,2},r_{3,1})$
& $(r_{3,1},r_{3,3})$ \\
\hline
$S_6$
& $(2r_{3,1},r_{3,2})$
& $(r_{3,1},r_{3,2})$
& $(r_{3,2},r_{3,4})$ \\
\hline
\end{tabular}
\caption{Queries for the $3K_2$ scheme.}
\label{tab:queries_3K2}
\end{table}

Now, once each server receives these queries, they apply the dot product for each of $r_{i,j}$ on its stored message. Let $\theta = 1$, then each server produces the following answers
\begin{align}
    A_{1}^{[1]} &= (a_1,c_1),\qquad \hspace{0.7em}    A_{2}^{[1]} = (b_1,d_1),\\
    A_{3}^{[1]} &= (\frac{1}{2}b_2,a_2),\qquad
A_{4}^{[1]} = (a_2,b_2),\\
    A_{5}^{[1]} &= (b_3,a_3), \qquad \hspace{0.7em}A_{6}^{[1]} = (2a_3,b_3).
\end{align}
The received classical bits upon measurement are given by
\begin{align}
    Y = M \begin{bmatrix}
        A_1^{[1]}\\
        A_2^{[1]}\\
        \vdots\\
        A_6^{[1]}
    \end{bmatrix} = \begin{bmatrix}
        a_1+a_2+b_3\\
        b_1+b_2+2a_3\\
c_1+\frac{1}{2} b_2+a_3\\
d_1+a_2+b_3\\
2d_1+a_2+b_3\\
c_1+b_2+2a_3
    \end{bmatrix}.
\end{align}
Now, to retrieve $a_1,b_1,c_1,d_1$, we apply the following linear decoding
\begin{align}
    a_1 &= y_1 - 2y_4+y_5,\\
    b_1& = y_2+2y_3-2y_6,\\
    c_1 &= 2y_3-y_6,\\
    d_1&= y_5-y_4.
\end{align}
In general, the decoding for each required message index is given in Table~\ref{tab:scalar_decoding_3K2}.
\begin{table}[h]
\centering
\begin{tabular}{|c|c|c|c|}
\hline
Coordinate & Request $W_1$ & Request $W_2$ & Request $W_3$ \\
\hline
$a_\theta$
& $y_1-2y_4+y_5$
& $y_3+y_6-y_2$
& $y_5+2y_4-2y_1$ \\
\hline
$b_\theta$
& $y_2+2y_3-2y_6$
& $y_4+y_1-y_5$
& $y_6+y_2-2y_3$ \\
\hline
$c_\theta$
& $2y_3-y_6$
& $2y_1-y_5$
& $y_2-y_3$ \\
\hline
$d_\theta$
& $y_5-y_4$
& $2y_6-y_2$
& $2y_4-y_1$ \\
\hline
\end{tabular}
\caption{Scalar decoding for the $3K_2$ scheme.}
\label{tab:scalar_decoding_3K2}
\end{table}
Since the scheme can retrieve $4$ symbols using 6 qudits, then the scheme achieves the rate $R_Q = \frac{4}{6}$.

The converse follows from the non-signaling principle and the cut-set bound \cite[Lemma 1]{quantum_dist_storage_lei}. Fix a message $W_i$, stored at servers $i,i+3$ and put these two servers storing $W_i$ on one side of the cut and the user along with other servers on the other side of the cut. 
Let $D_i$ be the number of downloaded qudits from the $i$th server. For decodability, we have
\begin{align}
    L \leq 2D_i + 2D_{i+3}.
\end{align}
Thus, summing over the three messages, we have
\begin{align}
    3L \leq 2 \sum_{i \in [6]}D_i,
\end{align}
showing that the rate is upper bounded as $R \leq \frac{2}{3}$ as required.

\bibliographystyle{unsrt}
\bibliography{references}

\end{document}